\documentclass[11pt]{article}
\usepackage[utf8]{inputenc}
\usepackage{amsfonts,amssymb,amsmath,enumerate}
\usepackage{theorem,cite}
\usepackage{indentfirst}
\usepackage{textcomp}
\usepackage{color}

\usepackage{latexsym}   
\usepackage{hyperref}
\usepackage{epsfig}
\usepackage{pgf,tikz}

\definecolor{dblue}{rgb}{0,0,0.7}
\hypersetup{colorlinks,citecolor=dblue,filecolor=black,linkcolor=dblue,urlcolor=blue}
\providecommand{\scr}{\mathcal}

\definecolor{dgreen}{rgb}{0,.55,0}

\theorembodyfont{\rmfamily}

\newcommand{\bea}{\begin{eqnarray}}
\newcommand{\eea}{\end{eqnarray}}
\newcommand{\beano}{\begin{eqnarray*}}
\newcommand{\eeano}{\end{eqnarray*}}
\newcommand{\beq}{\begin{equation}}
\newcommand{\eeq}{\end{equation}}
\newcommand{\nonu}{\nonumber \\}
\newcommand{\mb}[1]{\hspace{2.1ex}\mbox{#1}\hspace{2.1ex}}
\numberwithin{equation}{section}

\newcommand{\vph}{\varphi}

          \def\fc{{\mathfrak c}}
\def\fe{{\mathfrak e}}     \def\ff{{\mathfrak f}}
          
         \def\fo{{\mathfrak o}}
\def\fp{{\mathfrak p}}     \def\fq{{\mathfrak q}}         
             \def\fu{{\mathfrak u}}
\def\fz{{\mathfrak z}}

\newcommand{\II}{{\mathbb I}}

\newcommand{\OO}{{\mathbb O}}

\newcommand{\VV}{{\mathbb V}}

\def\ce{\langle\fe\rangle}
\def\cee{\langle\fe^2\rangle}
\def\cqq{\langle\bar \fq \fq\rangle}
\def\cqeq{\langle\bar \fq\fe \fq\rangle}
\def\cpp{\langle\bar \fp \fp\rangle}
\def\cpep{\langle\bar \fp\fe \fp\rangle}

\begin{document}
\title{\vspace{-6ex}\bf Finite $W$-superalgebras and
quadratic spacetime supersymmetries.   \vspace{-1ex}}
\author{%
    E. Ragoucy$^1$\footnote{ragoucy@lapth.cnrs.fr}, L.A.Yates$^2$\footnote{Luke.Yates@utas.edu.au} $\,$and P.D.Jarvis$^2$\footnote{Peter.Jarvis@utas.edu.au}
\\[1cm]
$^1$  LAPTh, CNRS and USMB, BP 110, F-74941 Annecy-le-Vieux Cedex, France\\
$^2$School of Natural Sciences, University of Tasmania, Hobart, Tasmania, Australia}
\maketitle

\abstract
We consider Lie superalgebras under constraints of Hamiltonian reduction, yielding finite $W$-superalgebras which provide candidates for quadratic spacetime superalgebras. These have an undeformed bosonic symmetry algebra (even generators) graded by a fermionic sector (supersymmetry generators) with anticommutator brackets which are quadratic in the even generators. We analyze the reduction of several Lie superalgebras of type $gl(M|N)$ or $osp(M|2N)$ at the classical (Poisson bracket) level, and also establish their quantum (Lie bracket) equivalents. Purely bosonic extensions are also considered. As a special case we recover a recently identified quadratic superconformal algebra, certain of whose unitary irreducible massless representations (in four dimensions) are ``zero-step" multiplets, with no attendant superpartners. Other cases studied include a six dimensional quadratic superconformal algebra with vectorial odd generators, and a variant quadratic superalgebra with  undeformed $osp(1|2N)$ singleton supersymmetry, and a triplet of spinorial supercharges.

\vspace{5mm}

\tableofcontents
\vfill
\pagebreak

\section{Introduction}
\label{sec:Introduction}
The analysis of quadratic algebras as spectrum generating algebras for the solution of quantum models\cite{sklyanin1985algebra}, or as a basis of generalized symmetry principles for physical systems, has led to a wide range of applications including special function theory \cite{zhedanov1991hidden} and
super-integrable systems \cite{kalnins2009models,genest2014racah}. Indeed, extensions of Lie algebra theory underlying exactly solvable models in mathematical physics, including Yangian and $W$-algebras, and related families of $q$-deformation, have been found to provide a rich source of examples of quadratic algebras and associated superalgebras. At the same time, substantial mathematical underpinnings have been developed for the structure of general quadratic algebras \cite{polishchuk2005quadratic}.

In this work we wish to introduce certain classes of (finite) quadratic $W$-superalgebras by Hamiltonian reduction from standard Lie superalgebras. 
The embeddings leading to the constraints are arranged in each case so as to leave undeformed, a Lie algebra or Lie superalgebra, together with conserved supercharges (supersymmetry) generators whose (anticommutator) brackets  
close on combinations quadratic in the even, undeformed generators. 

The motivation for this study is to provide constructions of finite 
$W$-superalgebras, which may in appropriate real forms provide alternative models of `supersymmetry' in nature. In particular, we 
analyze the reduction of several Lie superalgebras of type $gl(M|N)$ or $osp(M|2N)$ at the classical (Poisson bracket) level, to finite $W$-superalgebras, and also establish their quantum (Lie bracket) equivalents.
 Purely bosonic extensions are also considered. As a special case we recover by this method a recently identified quadratic conformal superalgebra (first obtained from a first principles construction \cite{yates2018hidden}), whose massless unitary irreducible representations (in four dimensions) have been shown to admit `zero step' multiplets, with no attendant superpartners. Other cases include a six dimensional quadratic conformal
superalgebra with $so(6)$ symmetry and vectorial supercharges, and a variant quadratic superalgebra with undeformed $osp(1|2N)$ singleton supersymmetry, with partners to the odd generators providing a triplet of spinorial supercharges. 

In section \ref{sec:FiniteW} below, we provide a brief resum\'{e} of the 
method of Hamiltonian reduction of a Poisson Lie (super) algebra, leading
to second class constraints solvable by Dirac brackets at the classical level,
yielding (in the quadratic case, via suitable symmetrization) to a quadratic
superalgebra at the quantum level (references to the literature are given below). Section \ref{sec:GlReductions} implements this 
for the superalgebra $gl(N|2)$\,, recovering (for $N=4$) the 
quadratic conformal superalgebra \cite{yates2018hidden} (as well as a purely bosonic equivalent based on $gl(N\!+\!2)$\,). The Casimir operators are also constructed for this case. Section \ref{sec:OspReductions} treats cases of reductions of orthosymplectic 
superalgebras $osp(M|2N)$, of type $(M|2)$ and $(3|2N)$, respectively.
Conclusions and further discussion are provided in section \ref{sec:Conclusion} .


\section{Finite $W$-algebras and superalgebras.}
\label{sec:FiniteW}
In this section we recall the basic method for the derivation of finite
$W$-algebras and superalgebras via symplectic reduction
(see \cite{MR1184379, MR1255424, MR1399720} and also \cite{MR3134037} for the theory of finite $W$-algebras and applications).
At the classical level, a finite-dimensional symmetry
superalgebra comprises a ${\mathbb Z}_2$-graded super Poisson-Lie
algebra of operators $A,B,\cdots$ with grading $[\,\mathbf{\cdot}\,]=0$ for even (bosonic)
and $[\,\mathbf{\cdot}\,]=1$ for odd (fermionic) generators, and Poisson brackets
that are graded antisymmetric
 and obey the graded Jacobi identity:
\beano
&&\{A\,,\,B\} = - (-1)^{[A][B]}\, \{B\,,\,A\}\\
&&\big\{A\,,\{B\,,\,C\}\big\} = \big\{\{A\,,\,B\}\,,\,C\big\} + (-1)^{[A][B]}\,\big\{B\,,\{A\,,\,C\}\big\}\,.
\eeano
In the presence of a system of second class constraints $\Phi\,$, with generators $\{\varphi_a, a=1,2,..\}$, a consistent Poisson bracket structure is built through the Dirac brackets, as follows. 
One first introduces the matrix $\Delta$ of pairwise Poisson brackets of all constraint generators: 
\[
\Delta_{ab}=\{\vph_a\,,\,\vph_b\}\,.
\]
It is invertible because the constraints are second class, and we define its inverse $\Delta^{-1}$ with entries $\Delta^{ab}$. Then the Dirac bracket is defined by
$$
\{A\,,\,B\}_* \simeq \{A\,,\,B\} - \sum_{a,b}\{A\,,\,\vph_a\}\, \Delta^{ab}\, \{\vph_b\,,\,B\}
$$
where the symbol $\simeq$ means that one has to apply the constraints \textit{once all Poisson brackets on the right hand side have been computed.}
The Dirac brackets are well-defined (graded antisymmetric, and obeying the graded Jacobi identity), and being consistent with the constraints in that
$
\{A\,,\,\vph\}_*=0\,,
$
for all operators $A$\,, and constraints $\vph\in\Phi$\,, effect a projection of the symplectic manifold which is the phase space of the system, on to the lower dimensional constraint surface.

For the cases to be treated, the matrix elements $\Delta_{ab}$ become scalars, and not operator (field) dependent, and the constraints are generators, whose Poisson brackets are linear. Hence the resultant $W$-algebra will have at most quadratic Dirac brackets. In this situation there is a simple way to quantize them, using symmetrization. It amounts to replacing all Dirac brackets by (anti-)commutators, and all products by their symmetrized version, for example $xy\,\to\, \frac12(xy+yx)$. Since the Jacobi identities are obeyed at the Poisson bracket level, the symmetrization ensures that they will still be obeyed at the quantum level. A final step is that 
the Casimirs of the $W$-algebra can be obtained via its embedding in the Lie (super)algebra, with appropriate use of the constraints.

In the standard technique, finite $W$-algebras and superalgebras are constructed
by analyzing various embeddings of $sl(2)$ in finite dimensional simple Lie algebras and superalgebras (for example, principal embeddings, in the
well-studied $W_n$ cases). They have been first introduced in a physics context, see e.g. \cite{MR1184379, MR1399720, MR1462770} and \cite{MR1965473} for the supersymmetric version, but then studied at the algebraic level by mathematicians, see for instance \cite{MR1255424, MR3134037, MR3147661}. The images of the $sl(2)$ generators encapsulate the second class constraints wherein the diagonal Cartan generator is constrained to vanish, and the positive root vector set to unity, leaving a reduced set of nonzero $W$ generators with quadratic Poisson bracket algebra. 
(Supersymmetric variants involving principal $osp(1|2)$ embeddings have also been considered). For the cases to be examined here, however, the $sl(2)$ embeddings are rather regular, and simply involve the identification of an appropriate isomorphic $sl(2)$ subalgebra carrying the constraints, augmented by the consistency requirement that certain of the odd generators should also be constrained to vanish. At the same time, a subalgebra of the even generators retains its standard Poisson brackets, so that the final quantum algebra generically becomes that of a Lie algebra graded by odd generators with quadratic anticommutator brackets -- and hence, in appopriate real forms,  a candidate `quadratic spacetime superalgebra' as described above.

\section{Quadratic superalgebras from $gl(N|2)$ reduction, and bosonic counterparts.}
\label{sec:GlReductions}

\subsection{The $W(gl(N|2),gl(2))$ superalgebras.}
The Lie superalgebra has generators $e_{ab}$, $1\leq a,b\leq N+2$, and the grading is defined by
$$
[e_{ab}]=[a]+[b] \mb{with} \begin{cases} [a]=0 \mb{for} 1\leq a\leq N\\
[a]=1 \mb{for} a= N+1,N+2\end{cases}\,.
$$
The Poisson brackets are given by
\beq\label{eq:relcom-superalg}
\{e_{ab},e_{cd}\} = \delta_{bc}\,e_{ad} - (-1)^{([a]+[b])([c]+[d])}\,\delta_{ad}\,e_{cb}\,.
\eeq
We construct the finite $W$-algebra associated to the constraints
\beq
\label{eq:const-1}
\begin{aligned}
&
\begin{aligned}
&e_{j,N+1}=\, 0\\
&e_{N+2,j}=\,0
\end{aligned}\qquad 1\leq j\leq N\mb{;}\\ 
&e_{N+2,N+1}=\,1\mb{;} \\
&e_{N+1,N+1}-e_{N+2,N+2}=\, 0\mb{.}
\end{aligned}
\eeq
The $W$-algebra is defined as the vector space generated by the unconstrained generators, and equipped with the Dirac brackets $\{\cdot,
\cdot \}_*$ associated to the above second class constraints. 

For the case under consideration, the $(2N\!+\!2)\times(2N\!+\!2)$ matrix $\Delta$ reads,
using the order $\vph_1=e_{1,N+1}$, $\vph_{N+1}=e_{N+2,1}$, etc.
(see \eqref{eq:const-1}),
\beq
\Delta=\begin{pmatrix}
\OO_N & \II_N & \fo_N & \fo_N \\ \II_N & \OO_N & \fo_N & \fo_N \\ 
(\fo_N)^t & (\fo_N)^t & 0 & 2 \\ (\fo_N)^t & (\fo_N)^t & -2 & 0 \end{pmatrix}
\eeq 
where $\OO_N$ (resp. $\II_N$) is a zero (resp. identity) square matrix of size $N$ and $\fo_N$ is a $N$-vector filled with zeros.
Then, it is a matter of calculation to get the Dirac brackets of the $W$-algebra. For ease of reading, we introduce
\beq
\begin{aligned}
&\fe_{ij}=e_{ij},\ 1\leq i,j\leq N \mb{;}\\ 
& \fu=\frac12(e_{N+1,N+1}+e_{N+2,N+2}) \mb{;}
\fz=e_{N+1,N+2}\mb{;}\\
&\fq_i=e_{i,N+2} \mb{;} \bar\fq_i=e_{N+1,i}\mb{;} \cqq 
=\sum_{i=1}^N \bar \fq_i \fq_i \mb{;}\\
& \ce=\sum_{i=1}^N e_{ii} \mb{;} \cee=\sum_{i,j=1}^N e_{ij}e_{ji} \mb{.}
\end{aligned}
\eeq
The Poisson brackets of the $W$-algebra read:
\beq
\begin{aligned}
&\{\fe_{ij}\,,\,\fe_{kl}\}_* = \delta_{kj}\,\fe_{il} - \delta_{il}\,\fe_{kj}
\mb{;}\\
& \{\fe_{ij}\,,\,\fu\}_* = 0 \mb{;} \{\fe_{ij}\,,\,\fz\}_* = 0 \mb{;} \{\fu\,,\,\fz\}_* = 0 \mb{;}\qquad \\ 
&\{\fe_{ij}\,,\,\fq_{k}\}_* = \delta_{kj}\,\fq_{i} 
\mb{;} \{\fe_{ij}\,,\,\bar\fq_{k}\}_* =  - \delta_{ik}\,\bar\fq_{j} \mb{;}\\
&\{\fu\,,\,\fq_{i}\}_* = -\frac12\,\fq_{i} 
\mb{;} \{\fu\,,\,\bar\fq_{i}\}_* =  \frac12\,\bar\fq_{i} \mb{;}\\
&\{\fz\,,\,\fq_{i}\}_* = \fu\,\fq_{i} + \sum_{k=1}^N \fe_{ik}\,\fq_{k}
\mb{;} \{\fz\,,\,\bar\fq_{i}\}_* =  -\big(\fu\,\bar\fq_{i} + \sum_{k=1}^N \bar\fq_{k}\fe_{ki}\big) \mb{;}\\
&\{\fq_i\,,\,\fq_{j}\}_* = 0 
\mb{;} \{\bar\fq_i\,,\,\bar\fq_{j}\}_* =  0 \\
&\{\fq_i\,,\,\bar\fq_{j}\}_* = \delta_{ij}\big(\fz-\fu^2\big) -2\,\fu\,\fe_{ij} - \sum_{k=1}^N \fe_{ik}\fe_{kj} \mb{.}
\end{aligned}
\label{fer}
\eeq
It is easy to see that 
\beq
\label{fer-first-cas}
\gamma_1=\fu+\frac12\ce\quad\mbox{and}\quad \gamma_2=\fz+\fu^2-\frac12\cee
\eeq
are central generators in the $W$-algebra. Using them to eliminate $\fu$ and $\fz$, we get 
\beq
\begin{aligned}
&\{\fe_{ij}\,,\,\fe_{kl}\}_* = \delta_{kj}\,\fe_{il} - \delta_{il}\,\fe_{kj} \mb{;}\\
&\{\fe_{ij}\,,\,\fq_{k}\}_* = \delta_{kj}\,\fq_{i} 
\mb{;} \{\fe_{ij}\,,\,\bar\fq_{k}\}_* =  - \delta_{il}\,\bar\fq_{j} \,;\\
&\{\fq_i\,,\,\fq_{j}\}_* = 0 
\mb{;} \{\bar\fq_i\,,\,\bar\fq_{j}\}_* =  0 \mb{;}\\\end{aligned}
\label{PB:eqqb}
\eeq
\beq
\{\fq_i\,,\,\bar\fq_{j}\}_* = \delta_{ij}\Big(\frac12\cee-\frac12\ce^2+\fc_1\ce+\fc_2\Big) -\big(\fc_1-\ce\big)\,\fe_{ij} - \sum_{k=1}^N \fe_{ik}\fe_{kj} \mb{,}\label{ser-w4}
\eeq
where $\fc_1=2\gamma_1$ and $\fc_2=\gamma_2-2\gamma_1^2$ are central generators.

In order to implement quantization, note that the two central elements $\gamma_1$ and $\gamma_2$ are already symmetrized. We can thus work\footnote{One can check easily that $\gamma_1$ and $\gamma_2$ are indeed central in the quantum/symmetrized version of the algebra \eqref{fer}.} at the level of relations \eqref{PB:eqqb}-\eqref{ser-w4}. The first two of these being linear, they remain unchanged (except from the change from Poisson brackets to commutators or anti-commutators). For the last one, one has to symmetrize the two products
\bea
&& \langle\fe\rangle\fe_{ij}\ \to\ \frac12\big( \langle\fe\rangle\fe_{ij} + \fe_{ij}\langle\fe\rangle \big) = \langle\fe\rangle\fe_{ij} \mb{;}
 \\
&& \sum_{k=1}^N \fe_{ik}\fe_{kj}\ \to\ \frac12\big(\sum_{k=1}^N \fe_{ik}\fe_{kj} + \sum_{k=1}^N \fe_{kj}\fe_{ik}\big)=\sum_{k=1}^N \fe_{ik}\fe_{kj}  +\frac12\,\big( \delta_{ij}\langle\fe\rangle-N\, \fe_{ij}\big)\,,
\eea
which leads to the anti-commutator 
\beq
[\fq_i\,,\,\bar\fq_{j}]_+ = \delta_{ij}\Big(\frac12\cee-\frac12\ce^2+(\fc_1-\frac12)\ce+\fc_2\Big)
 -\big(\fc_1-\frac{N}{2}-\ce\big)\,\fe_{ij} - \sum_{k=1}^N \fe_{ik}\fe_{kj}\,.
\label{Wgl-anticomm}
\eeq
Together with the standard commutations of $\fq_i\,,\,\bar\fq_{j}$ and
$\fe_{ij}$ (see eqs \eqref{PB:eqqb}), this superalgebra is precisely that of \cite{yates2018hidden}, denoted $gl_2(N/1)^{\alpha,c}$ (see \cite{yates2018hidden} equation (6)),
which was obtained by a first principles construction, and investigated as a potential quadratic spacetime supersymmetry algebra (for further discussion see section \ref{sec:Conclusion} below). Explicitly, the correspondence is given by
\beq
{E^i}_j\ \to\ -e_{ji}\ ;\ Q^i\ \to\ \fq_i\ ;\ \bar Q^i\ \to\ -\bar\fq_i\ ;\ \alpha\ \to\ -(\fc_1+\frac{N}2)\ ;\ 
c\ \to\ \fc_2\,.
\eeq
Note that the $W$-algebra framework indicates that the parameters $\alpha$ and $c$ should be considered as central generators, a fact that could be of importance in the study of the representations of the superalgebra.

To extract the central elements for this $W$-superalgebra, one considers the $(N\!+\!2)\times(N\!+\!2)$ Gel'fand  matrix of reduced generators, derived from the original superalgebra with constraints imposed, 
\beq
\label{eq:Ematrix}
E=\begin{pmatrix} \fe^{11} & \cdots& \fe^{1N} & 0 & \fq^{1} \\
\vdots & & \vdots & \vdots & \vdots \\
 \fe^{N1} & \cdots & \fe^{NN} & 0 & \fq^{N} \\ \bar\fq^{1} & \cdots & \bar\fq^{N} & \fu & \fz \\
 0 & \cdots & 0 & 1 & \fu \end{pmatrix}\,.
\eeq
The Casimir operators $\gamma_i\,, i=0,1,2,\cdots$\,, 
are traces of appropriately graded matrix powers of $E$
(explicit expressions for $gl(M|N)$ are given 
in \cite{jarvis1979casimir})\footnote{A generating function for them is given by the expansion of the superdeterminant (Berezinian), 
$Ber(E-x\,\II_{N+2})$\,,
with $\II_{N+2}$ the identity matrix \cite{Nazarov1991aa, 10.1155/S1073792804132935}.
It is related to a supersymmetric version of the Capelli identity.}. Note that since the $W$-algebra is not linear, this result applies only to the classical version. Lower degree terms can occur in the quantum version, see below.
Direct calculation for the lowest degrees shows that $\gamma_1$ and 
$\gamma_2$, up to a sign and a factor of $\frac12$, have the expression given in \eqref{fer-first-cas} (with $\gamma_0=1$). Their quantum versions remain unchanged.
At degrees 3 and 4 we find\footnote{We omit polynomials in $\fc_1$ and $\fc_2$, since they are already central at the quantum level.}
\beq
\begin{aligned}
\gamma_3 &=\langle \fe^3\rangle - 3 \cqq + 2 \fu^3 + 6 \fu\fz\,,\\
\gamma_4 &= \langle \fe^4\rangle - 4 \cqeq +8\fu\cqq -2\fz^2 + 12 \fz\fu^2 -2 \fu^4\,,
\end{aligned}
\eeq
where again $\fu$ and $\fz$ may be rewritten in terms of $\gamma_1$ and $\gamma_2$ as above. 
As already stated, $\gamma_3$ (resp. $\gamma_4$) is the classical Casimir of degree 3 (resp. degree 4), i.e. it has vanishing
Poisson brackets with any element in the (classical version of the) $W$-superalgebra. There are quantum corrections to these classical versions, and we get for the third and fourth Casimir of the quantum $W$-superalgebra:
\beq\label{eq:c34-uz}
\begin{aligned}
\fc_3&=\gamma_3+\frac{N-6}2\fz -\frac{5(N-2)}2\fu^2 
-5\fc_1\fu -\frac{(N-1)(N-2)}2\fu
\,,\\
 \fc_4&=
 \gamma_4+(N-6)\cqq +6(N-2)\fu^3+2(N+2)\fu\fz +4\fc_1(\fz+2\fu^2)+ 
 (4\fc_2+2\fc_1^2)\fu\\
 &\quad
 +\frac{N^2-8}2\fz-\frac{(N-2)(N+6)}2\fu^2 
-(3N+4)\fc_1\fu-\frac{(N-1)(N^2-4)}2\fu.
\end{aligned}
\eeq
Expressions \eqref{eq:c34-uz} were obtained by direct calculation, demanding that $\fc_3$ and $\fc_4$ are indeed central in the superalgebra \eqref{Wgl-anticomm}. We checked that the additional next-to-leading order terms in $\fc_3$ exactly correspond to the symmetrization of $\gamma_3$. This is not enough to get the quantum version of $\gamma_3$, since we are considering expressions which are not quadratic anymore. The last term, which is a next-to-next-to-leading order term, has to be found by brute calculation. Similarly, the first line in the expression of $\fc_4$ is obtained through symmetrization. When $N=4$, which corresponds to a quadratic superconformal algebra, we conjecture they are the only Casimir operators of this algebra.

As above, one can eliminate $\fu$ and $\fz$ using $\fc_1$ and $\fc_2$. 
For instance, $\fc_3$ can be rewritten as
\beq\label{eq:c3-e}
\begin{aligned}
\fc_3 &= \langle \fe^3\rangle - 3 \cqq -\frac32 \langle \fe^2\rangle\langle \fe\rangle
+\frac12\langle \fe\rangle^3 +\frac{6\fc_1+N-6}4\langle \fe^2\rangle
-\frac{6\fc_1+3N-8}4\langle \fe\rangle^2\\
&\quad+\frac{6(N-1)\fc_1-12\fc_2+(N-1)(N-2)}4\langle \fe\rangle\\
&\quad+3\fc_1\fc_2+\fc_1^3 +\frac{N-6}2\fc_2 -\frac{N+4}2\fc_1^2 -\frac{(N-1)(N-2)}4\fc_1,
\end{aligned}
\eeq
where the last line can be dropped out, since it is central on its own.

\subsection{The $W(gl(N+2),gl(2))$ algebras.}
We can construct the `bosonic version' of the finite $W$-algebra defined above. We start with the $gl(N\!+\!2)$ algebra, with Poisson brackets
\beq\label{eq:relcom-alg}
\{e_{ij},e_{kl}\} = \delta_{kj}\,e_{il} - \delta_{il}\,e_{kj}
\eeq
and consider the constraints
\beq
\label{eq:bos-const}
\begin{aligned}
&
\begin{aligned}
&e_{j,N+1}=\, 0\\
&e_{N+2,j}=\,0
\end{aligned}\qquad 1\leq j\leq N\mb{;}\\ 
&e_{N+2,N+1}=\,1\mb{;} \\
&e_{N+1,N+1}-e_{N+2,N+2}=\, 0\mb{.}
\end{aligned}
\eeq
The  $(2N\!+\!2)\times(2N\!+\!2)$ matrix $\Delta$ reads,
(using the order given in \eqref{eq:bos-const}, viz. $\vph_1=e_{1,N+1}$, $\vph_{N+1}=e_{N+2,1}$, etc.,)
\beq
\Delta=\begin{pmatrix}
\OO_N & \II_N & \fo_N & \fo_N \\ -\II_N & \OO_N & \fo_N & \fo_N \\ 
(\fo_N)^t & (\fo_N)^t & 0 & 2 \\ (\fo_N)^t & (\fo_N)^t & -2 & 0 \end{pmatrix}
\eeq 
where $\OO_N$ (resp. $\II_N$) is a zero (resp. identity) square matrix of size $N$ and $\fo_N$ is a $N$-vector filled with zeros.

Again, we introduce
\beq
\begin{aligned}
&\fe_{ij}=e_{ij},\ 1\leq i,j\leq N \mb{;}\\ 
& \fu=\frac12(e_{N+1,N+1}+e_{N+2,N+2}) \mb{;} 
\fz=e_{N+1,N+2}\mb{;}\\
&\fp_i=e_{i,N+2} \mb{;} \bar\fp_i=e_{N+1,i}\mb{;}\\
& \ce=\sum_{i=1}^N e_{ii} \mb{;} \cee=\sum_{i,j=1}^N e_{ij}e_{ji} \mb{.}
\end{aligned}
\eeq
The Poisson brackets of the $W$-algebra read:
\beq
\begin{aligned}
&\{\fe_{ij}\,,\,\fe_{kl}\}_* = \delta_{kj}\,\fe_{il} - \delta_{il}\,\fe_{kj}
\mb{;} \\
& \{\fe_{ij}\,,\,\fu\}_* = 0 \mb{;} \{\fe_{ij}\,,\,\fz\}_* = 0 \mb{;} \{\fu\,,\,\fz\}_* = 0 \mb{;} \\
&\{\fe_{ij}\,,\,\fp_{k}\}_* = \delta_{kj}\,\fp_{i} 
\mb{;} \{\fe_{ij}\,,\,\bar\fp_{k}\}_* =  - \delta_{il}\,\bar\fp_{j}\mb{;}  \\
&\{\fu\,,\,\fp_{i}\}_* = -\frac12\,\fp_{i} 
\mb{;} \{\fu\,,\,\bar\fp_{i}\}_* =  \frac12\,\bar\fp_{i} \mb{;} \\
&\{\fz\,,\,\fp_{i}\}_* = \fu\,\fp_{i} - \sum_{k=1}^N \fe_{ik}\,\fp_{k}
\mb{;} \{\fz\,,\,\bar\fp_{i}\}_* =  -\big(\fu\,\bar\fp_{i} - \sum_{k=1}^N \bar\fp_{k}\fe_{ki}\big)\\
&\{\fp_i\,,\,\fp_{j}\}_* = 0 
\mb{;} \{\bar\fp_i\,,\,\bar\fp_{j}\}_* =  0 \mb{;} \\
&\{\fp_i\,,\,\bar\fp_{j}\}_* = \delta_{ij}\big(\fu^2-\fz\big) -2\,\fu\,\fe_{ij} + \sum_{k=1}^N \fe_{ik}\fe_{kj}\mb{.} 
\end{aligned}
\eeq
It is easy to see that 
\beq\label{first-cas}
\gamma_1=\fu+\frac12\ce \mb{and} \gamma_2=\fz+\fu^2+\frac12\cee
\eeq
 are central generators in the $W$-algebra. Using them to eliminate $\fu$ and $\fz$, we get 
\beq
\begin{aligned}
&\{\fe_{ij}\,,\,\fe_{kl}\}_* = \delta_{kj}\,\fe_{il} - \delta_{il}\,\fe_{kj} \mb{;} 
\\
&\{\fe_{ij}\,,\,\fp_{k}\}_* = \delta_{kj}\,\fp_{i} 
\mb{;} \{\fe_{ij}\,,\,\bar\fp_{k}\}_* =  - \delta_{il}\,\bar\fp_{j} \mb{;}\\
&\{\fp_i\,,\,\fp_{j}\}_* = 0 
\mb{;} \{\bar\fp_i\,,\,\bar\fp_{j}\}_* =  0\, \mb{;} \\
&\{\fp_i\,,\,\bar\fp_{j}\}_* = \delta_{ij}\Big(\frac12\cee+\frac12\ce^2-\fc_1\ce-\fc_2\Big)
 -\big(\fc_1-\ce\big)\,\fe_{ij} + \sum_{k=1}^N \fe_{ik}\fe_{kj}
\end{aligned}
\eeq
where $\fc_1=2\gamma_1$ and $\fc_2=\gamma_2-2\gamma_1^2$ are central generators.

Quantization of the quadratic $W$-algebra follows the same procedure as with the fermionic counterpart above, leading to the commutator
$$
[\fp_i\,,\,\bar\fp_{j}] = \delta_{ij}\Big(\frac12\cee+\frac12\ce^2-(\fc_1-\frac12)\ce-\fc_2\Big)
 -\big(\fc_1+\frac{N}{2}-\ce\big)\,\fe_{ij} + \sum_{k=1}^N \fe_{ik}\fe_{kj}\,.
$$

The Casimir operators of this bosonic $W$-algebra can be extracted in the same way as for the fermionic counterpart above, using now the 
traces of standard matrix powers of the Gel'fand generators with constraints (or defining the appropriate generating function using the determinant instead of the Berezinian). 
We have already defined the two first Casimir operators $\fc_1$ and $\fc_2$.
At degrees 3 and 4, we find for the classical Casimir operators
\beq
\begin{aligned}
\gamma_3 &=\langle \fe^3\rangle + 3 \cpp + 2 \fu^3 + 6 \fu\fz
\\
\gamma_4 &=\langle \fe^4\rangle + 4 \cpep + 8\fu \cpp + 2\fz^2
+12\fz \fu^2 + 2 \fu^4 
\end{aligned}
\eeq
and for their quantum analogs
\beq
\begin{aligned}
\fc_3&=\gamma_3 
 -\frac{N+6}2\fz +\frac{5(N+2)}2\fu^2 -5\fc_1\,\fu-\frac{(N+2)(N+1)}2\,\fu\,,\\
 \fc_4&=
 \gamma_4-(N+6)\cqq +6(N+2)\fu^3+2(N-2)\fu\fz -4\fc_1(\fz+2\fu^2)- 
 (4\fc_2+2\fc_1^2)\fu\\
 &\quad
 -\frac{N^2-8}2\fz+\frac{(N+2)(N-6)}2\fu^2 
-(3N-4)\fc_1\fu-\frac{(N-1)(N^2-4)}2\fu.
\end{aligned}
\eeq


\section{Quadratic superalgebras from $osp(M|2N)$ reductions.}
\label{sec:OspReductions}
The procedure we have described so far in the context of  $gl(M|N)$  can be applied to other superalgebras. In this section we examine orthosymplectic cases, leading to the construction of quadratic superalgebras with $so(N)$ or $sp(2N)$ bosonic subalgebras. 

The $osp(M|2N)$ Lie superalgebra can be constructed as the folding of the $gl(M|2N)$ algebra\cite{frappat1996dictionary,jarvis1983casimir}. Recall that the $gl(M|2N)$ superalgebra has Poisson
brackets 
\beq
\{e_{ab},e_{cd}\} = \delta_{bc}\,e_{ad} - (-1)^{([a]+[b])([c]+[d])}\,\delta_{ad}\,e_{cb}
\,,\qquad 1\leq a,b,c,d\leq M+2N
\eeq
 and grading
\beq
[a]=\begin{cases} 0 \mb{for} 1\leq a\leq M\\ 1 \mb{for} M< a\leq M+2N
\end{cases}
\eeq
The superalgebra admits a morphism $\vph$\,, defined as
\beq\label{eq:phi}
\vph:\quad\begin{cases} gl(M|2N) & \to \quad gl(M|2N) \\ 
e_{ab} & \to \quad -(-1)^{[a]([b]+1)}\,\theta_a\theta_b\,e_{\bar b\bar a}\mb{,} 
\end{cases}
\eeq
where 
\beq
\begin{aligned}
&\theta_a=1 &\quad & \bar a=M+1-a &\quad & 1\leq a \leq M \mb{;} \\
&\theta_a=sg\big(M+N+\frac12-a) &\quad & \bar a=2M+2N+1-a &\quad & M+1\leq a \leq M+2N \mb{.} 
\end{aligned}
\eeq
$\vph$ is a Poisson algebra morphism, viz. $\vph\big(\{A,B\}\big) = \big\{\vph(A),\vph(B)\big\}$.
Note the property $[a]=[\bar a]$ and $(-1)^{[a]}\theta_a\theta_{\bar a}=1$, for all $a$, which shows that $\vph$ is an involution, $\vph^2=id$. Then the $gl(M|2N)$ superalgebra can be decomposed as a sum of the two $\vph$-eigenspaces.
The $osp(M|2n)$ algebra can be viewed as the $\vph$-eigenspace corresponding to the eigenvalue $+1$. Explicitly, the generators are constructed as
\beq\label{eq:ops-gen}
s_{ab}=e_{ab} -(-1)^{[a]([b]+1)}\,\theta_a\theta_b\,e_{\bar b\bar a}
= -(-1)^{[a]([b]+1)}\,\theta_a\theta_b\,s_{\bar b\bar a}\mb{,}
\eeq
leading to the Poisson brackets
\bea
\{s_{ab},s_{cd}\} &=& \delta_{bc}\,s_{ad} - (-1)^{([a]+[b])([c]+[d])}\,\delta_{ad}\,s_{cb}
\nonu
&&+(-1)^{([a]+[b])([c]+1)}\,\theta_a\theta_b\,\delta_{d\bar b}\,s_{c\bar a}
-(-1)^{[a]([b]+1)}\,\theta_a\theta_b\,\delta_{a\bar c}\,s_{\bar bd}\mb{.}
\eea
For our purpose, we specialize the notation to two particular cases, that we now describe in detail.

\subsection{The $W(osp(N|2),sp(2))$ superalgebras.}
As bosonic generators, one can choose $f_{ij}=e_{ij}-e_{\bar\jmath\bar\imath}$, $1\leq {i,j}\leq N$, 
with $\bar\imath=N+1-i$, together with 
$e_+=e_{N+1,N+2}$, $e_-=e_{N+2,N+1}$, $e_0=e_{N+1,N+1}-e_{N+2,N+2}$. The generators
$f_{ij}$ form an $so(N)$ algebra while $e_\mu$, $\mu=0,\pm$ generate an $sp(2)$ algebra. 
The Poisson brackets read
\bea 
\nonumber
\{f_{ij},f_{kl}\} &=& \delta_{kj}\,f_{il} - \,\delta_{il}\,f_{kj} 
+\delta_{l\bar \jmath}\,f_{k\bar \imath}-\delta_{i\bar k}\,f_{\bar \jmath l}
\mb{with} f_{\bar\jmath\bar\imath}=-f_{ij}\mb{;}\\  
\{e_0,e_\pm\} &=& \pm2\, e_\pm \mb{,} \{e_+,e_-\}=e_0\mb{;} 
\\
\{f_{ij},e_{\mu}\} &=&0,\qquad 1\leq i,j\leq N\,,\quad \mu=0,\pm \mb{.} 
\nonumber
\eea
The fermionic generators are $q_i^+=e_{i,N+2}+e_{N+1,\bar\imath}$ and 
$q_i^-=e_{i,N+1}-e_{N+2,\bar\imath}$, with Poisson brackets
\beq
\begin{aligned}
\{f_{ij},q^\pm_{k}\} &= \delta_{kj}\,q^\pm_{i} -\delta_{\bar\imath k}\,q^\pm_{\bar \jmath }\,,
&\qquad& \{e_0,q^\pm_{k}\} = \pm\,q^\pm_{k}\,,
\\
 \{e_\pm,q^\pm_{k}\} &=0\,,
&& \{e_\pm,q^\mp_{k}\}=-q^\pm_{k}\,,
\\
\{q^\pm_{i},q^\pm_{j}\} &=\pm2\,\delta_{i\bar\jmath}\,e_\pm\,,
&& 
\{q^+_{i},q^-_{j}\} =\delta_{i\bar\jmath}\,e_0-f_{\bar\jmath,i}\,.
\end{aligned}
\eeq
Note that in this case all $\theta$'s are 1, except $\theta_{N+2}=-1$.
We impose the constraints 
\beq
q^-_i=0 \mb{;} e_-=1\mb{;} e_0=0 \mb{;}
\eeq
leading to a matrix 
\beq
\Delta=\begin{pmatrix} -2\,\VV_{N} & \fo_N & \fo_N \\
 (\fo_N)^t & 0 & 2\\  (\fo_N)^t  & -2 & 0
 \end{pmatrix}
 \mb{with} \VV_N=\begin{pmatrix} 0  & \dots  & 0 & 1 \\ \vdots &&  & 0\\
0& 1& &\vdots \\ 1&0 &\dots &0\end{pmatrix}.
\eeq 
The Dirac brackets associated to these constraints define the $W(osp(N|2),sp(2))$ super-algebra.
It has generators
\bea
&& \ff_{ij}=f_{ij}\mb{;} \fz = e_+=e_{N+1,N+2}\mb{;} \fq_i=q_i^+=e_{i,N+2}+e_{N+1,\bar\imath}\mb{,}
\eea
 with Poisson brackets:
\beq\label{wospn2-PB} 
\begin{aligned}
\{\ff_{ij},\ff_{kl}\}_* &= \delta_{kj}\,\ff_{il} - \,\delta_{il}\,\ff_{kj} 
+\delta_{l\bar \jmath}\,\ff_{k\bar \imath}-\delta_{i\bar k}\,\ff_{\bar \jmath l}
\mb{with} \ff_{\bar\jmath\bar\imath}=-\ff_{ij}\mb{;}\\
\{\ff_{ij},\fq_{k}\}_* &= \delta_{kj}\,\fq_{i} -\delta_{\bar\imath k}\,\fq_{\bar \jmath }\mb{;}
\\
\{\fz,\fq_{i}\}_* &= \frac12\sum_{k=1}^N\ff_{ik} \,\fq_{k} \mb{;}
\{\fz,\ff_{ij}\}_*=0\mb{;}
\\
\{\fq_i,\fq_j\}_* &= 2\,\delta_{i,\bar\jmath}\,\fz -\frac12\,\sum_{k=1}^N\ff_{ik}\ff_{k\bar\jmath}\mb{.}
\end{aligned}
\eeq
Similarly to the $gl(N|2)$ case, the generator 
\beq
\gamma_2=\fz-\frac18\langle\ff^2\rangle=\fz-\frac18\sum_{k,j=1}^N \ff_{jk}\ff_{kj}
\eeq
is central, and we use it to eliminate $\fz$. It leads to the final form for the Poisson brackets of the $W$-superalgebra:

\beq\label{wospn2-PB2} 
\begin{aligned}
\{\ff_{ij},\ff_{kl}\}_* &= \delta_{kj}\,\ff_{il} - \,\delta_{il}\,\ff_{kj} 
+\delta_{l\bar \jmath}\,\ff_{k\bar \imath}-\delta_{i\bar k}\,\ff_{\bar \jmath l}
\mb{with} \ff_{\bar\jmath\bar\imath}=-\ff_{ij}\mb{;}\\
\{\ff_{ij},\fq_{k}\}_* &= \delta_{kj}\,\fq_{i} -\delta_{\bar\imath k}\,\fq_{\bar \jmath }\mb{;}
\\
\{\fq_i,\fq_j\}_* &= 2\,\delta_{i,\bar\jmath}\,\big(\gamma_2+\frac18\langle\ff^2\rangle\big) -\frac12\,\sum_{k=1}^N\ff_{ik}\ff_{k\bar\jmath}\mb{.}
\end{aligned}
\eeq

We perform quantization through symmetrization. In the quantization  of the algebra \eqref{wospn2-PB}, 
 two terms have to be symmetrized: $\sum_{k=1}^N\ff_{ik}\ff_{k\bar\jmath}$ and 
$\sum_{k=1}^N\ff_{ik} \,\fq_{k}$. We obtain
\beq
\begin{aligned}
{}[\ff_{ij},\ff_{kl}] &= \delta_{kj}\,\ff_{il} - \,\delta_{il}\,\ff_{kj} 
+\delta_{l\bar \jmath}\,\ff_{k\bar \imath}-\delta_{i\bar k}\,\ff_{\bar \jmath l}
\mb{with} \ff_{\bar\jmath\bar\imath}=-\ff_{ij}\mb{;}\\
{[\ff_{ij},\fq_{k}]} &= \delta_{kj}\,\fq_{i} -\delta_{\bar\imath k}\,\fq_{\bar \jmath }\mb{;}
\\
{[}\fq_i,\fq_j]_+ &= 2\,\delta_{i,\bar\jmath}\,\fz -\frac12\,\sum_{k=1}^N\ff_{ik}\ff_{k\bar\jmath} +\frac{N-2}4\,\ff_{i\bar\jmath}\mb{;}\\
{[\fz,\fq_j]} &= \frac12\sum_{k=1}^N\ff_{jk} \,\fq_{k} -\frac{N-2}4\,\fq_j\,.
\end{aligned}
\eeq
One can check that $\fc_2=\fz-\frac18\langle\ff^2\rangle$ is still central, and we can use it to eliminate $\fz$ at the quantum level. Of course, the resulting algebra is the same as the quantization through symmetrization of \eqref{wospn2-PB2}.

The superalgebra contains as bosonic part, an $so(N)$ algebra, and an $N$-vector of fermions $\fq^i$\,, that close quadratically on the orthogonal subalgebra. It should be interesting to look at  a non-compact version of the case $N=6$, corresponding to the conformal $so(4,2)$ algebra.

\subsection{The $W(osp(3|2N),so(3))$ superalgebras.}
The bosonic generators now are $e^+=e_{12}-e_{23}$,
$e^-=e_{21}-e_{32}$, $e^0=e_{11}-e_{33}$ which form an $so(3)$ subalgebra, together with
$f_{ij}=e_{i+3,j+3}-\theta_i\theta_j\,e_{3+\bar\jmath,3+\bar\imath}$ (with $\bar\imath=2n+1-i$) which generate $sp(2N)$\,,
and where we have redefined
$$
\theta_j=\begin{cases} +1\,,\quad 1\leq j\leq N\\ -1 \,,\quad N+1\leq j\leq 2n\,.
\end{cases}
$$
 Then the Poisson brackets read
\bea 
\nonumber
\{f_{ij},f_{kl}\} &=& \delta_{kj}\,f_{il} - \,\delta_{il}\,f_{kj} 
+\theta_i\theta_j\,\big(\delta_{l\bar \jmath}\,f_{k\bar \imath}-\delta_{i\bar k}\,f_{\bar \jmath l}\big)
\mb{with} f_{\bar\jmath\bar\imath}=-\theta_i\theta_j\,f_{ij}\mb{;}\\
\{e_0,e_\pm\} &=& \pm\, e_\pm \mb{;} \{e_+,e_-\}=e_0\mb{;}
\\
\{f_{ij},e_{\mu}\} &=&0,\qquad 1\leq i,j\leq N\,,\quad \mu=0,\pm\mb{.}
\nonumber
\eea
The fermionic generators take now the form $q_i^+=e_{1,i+3}-\theta_i\,e_{3+\bar\imath,3}$,
$q_i^0=e_{2,i+3}-\theta_i\,e_{3+\bar\imath,2}$ and 
$q_i^-=e_{3,i+3}-\theta_i\,e_{3+\bar\imath,1}$, with Poisson brackets
\beq
\begin{aligned}
\{f_{ij},q^\mu_{k}\} &= \theta_i\theta_j\,\big(-\delta_{ik}\,q^\mu_{j} +\delta_{\bar\jmath k}\,q^\mu_{\bar \imath}\big) &\quad&
\{e_0,q^\mu_{k}\} = \mu\,q^\mu_{k} && \mu=0,\pm \mb{;}\\
 \{e_\pm,q^\pm_{k}\}&=0\mb{;}
&& \{e_\pm,q^\mp_{k}\}=\mp\,q^0_{k}\mb{;}
&& \{e_\pm,q^0_{k}\}=\pm\,q^\pm_{k}\mb{;}
\\
\{q^\pm_{i},q^\pm_{j}\} &=0\mb{;}
&& 
\{q^0_{i},q^0_{j}\} =-\theta_i\,f_{\bar\imath j}\mb{;}&&
\\
\{q^+_{i},q^-_{j}\} &=\theta_i\big(\delta_{i\bar\jmath}\,e_0-f_{\bar\imath j}\big)\mb{;}
&& 
\{q^0_{i},q^\pm_{j}\} =\theta_i\,\delta_{i\bar\jmath}\,e_\pm\mb{.} &&
\end{aligned}
\eeq
We impose the constraints $e_-=1$ and $e_0=0$, leading the the Dirac brackets
\beq
\begin{aligned}
\{\ff_{ij},\ff_{kl}\}_* &= \delta_{kj}\,\ff_{il} - \,\delta_{il}\,\ff_{kj} 
+\theta_i\theta_j\,\big(\delta_{l\bar \jmath}\,\ff_{k\bar \imath}-\delta_{i\bar k}\,\ff_{\bar \jmath l}\big)
\mb{with} \ff_{\bar\jmath\bar\imath}=-\theta_i\theta_j\,\ff_{ij}\mb{;}\\
\{\ff_{ij},\fe_{+}\}_* &=0,\qquad 1\leq i,j\leq 2n \mb{;}
\end{aligned}
\eeq
\beq
\begin{aligned}
\{\ff_{ij},\fq^\mu_{k}\}_* &= \theta_i\theta_j\,\big( \delta_{\bar\jmath k}\,\fq^\mu_{\bar \imath} -\delta_{ik}\,\fq^\mu_{j} \big) \qquad \mu=0,\pm\mb{;}\\
\{\fe_+,\fq^+_{k}\}_* &= \fe^+\,\fq^0_{k}\,,  \qquad
 \{\fe_+,\fq^0_{k}\}_* =\fq^+_{k}-\fe^+\,\fq^-_{k}\,,
\qquad \{\fe_+,\fq^-_{k}\}_*=-\fq^0_{k}\mb{;}
\end{aligned}
\eeq
\beq
\begin{aligned}
\{\fq^+_{i},\fq^+_{j}\}_* &= \fq^+_{i}\,\fq^0_{j}- \fq^0_{i}\,\fq^+_{j} \,,
&\quad& 
\{\fq^+_{i},\fq^0_{j}\}_* = -\fq^+_{i}\,\fq^-_{j}+\theta_j\,\delta_{\bar\imath j}\,\fe_+\,,
\\
\{\fq^+_{i},\fq^-_{j}\}_* &=\fq^0_i\fq^-_j- \theta_i\,\ff_{\bar\imath j}\,,
&& 
\{\fq^0_{i},\fq^-_{j}\}_* =-\fq^-_i\fq^-_j-\theta_j\,\delta_{i\bar\jmath} \,,
\\
\{\fq^0_{i},\fq^0_{j}\}_* &= \theta_i\,\ff_{\bar\jmath i}\,,
&& 
\{\fq^-_{i},\fq^-_{j}\}_* =0\,.
\end{aligned}
\eeq
The superalgebra is quadratic in fermions, and thus is a variant on the 
standard construction from $W(gl(N\!+\!2|2),gl(2))$ in section 
\ref{sec:GlReductions} above. Moreover, it contains an undeformed $osp(1|2N)$ singleton superalgebra (generated by $\fq^0_{i}$ and $\ff_{ij}$)\,, and so in real forms is a candidate for higher dimensional quadratic conformal spacetime supersymmetry.
We leave the quantization step to be completed by (anti)symmetrization, as in the cases already examined above.


\section{Conclusion}
\label{sec:Conclusion}
In this paper we have provided explicit constructions for a class
of candidate quadratic spacetime supersymmetries as $W$-superalgebras, via
systematic application of Hamiltonian reduction to certain simple
Lie superalgebras. 

In particular we recover 
as $W(gl(N\!+\!2|2), gl(2))$, the two-parameter family of quadratic superalgebras analysed in \cite{yates2018hidden}\, there denoted $gl_2(N/1)^{\alpha, c}$\, (see also \cite{jarvis_2011,jarvis_2012}).
These were first obtained from a first principles construction, and 
their consistency as algebras of PBW type established via the formal theory of 
abstract quadratic algebras \cite{polishchuk2005quadratic}. For the $N=4$ real form with even part $u(2,2) \cong so(4,2) +u(1)$, a Kac module-type
construction led to the observation that for certain parameter choices,
there are `zero step' unitary irreducible representations which 
comprise a single module of the even subalgebra, namely, a multiplet corresponding to one of the standard  physical massless conformal fields of spin $0$, $\frac 12$ or $1$, with the odd generators being identically zero (the standard conformal supersymmetry
Lie superalgebra is a contraction limit of $gl_2(N/1)^{\alpha, c}$). This scenario of `ultra-short supermultiplets' can thus be likened to an extreme form of \emph{unbroken} supersymmetry, where a (degenerate) ground state is annihilated by the supercharges, but with \emph{no} other (paired) states present in the spectrum. This scenario has potential implications for phenomenology in the context of particle symmetries, and the absence of super-partners.

With the identification of this and other potential candidate quadratic spacetime supersymmetry algebras as $W$-superalgebras, which
have been intensively studied in recent years,
formal tools from that work are available for their investigation. These include direct results on Casimir operators (as discussed above), and also methods for constructing representations, and providing a more general theory for the existence of special `zero step', no-superpartner cases. On this point, the tuning of the parameters identifying which of the family of quadratic superalgebras, admit the existence of such `hyper-atypical' representations \cite{yates2018hidden}, can be seen instead, in the context of the $W$ construction, as the selection of representations of the primary superalgebra (which is to undergo Hamiltonian reduction), which have fixed values of the lowest Casimir invariants.

In this work we have also used the Hamiltonian reduction formlism to construct new finite quadratic $W$-superalgebras, beyond the above superconformal case, which are equally candidates for quadratic spacetime superalgebras. If appropriate real forms exist, their unitary representations may supply a tool kit of interesting variants on (quadratic) `extended supersymmetries'.
This applies to the $W$-superalgebras derived from $osp(M|2N)$ reductions
in section \ref{sec:OspReductions} above, which include a case with 6 dimensional orthogonal (conformal) group as bosonic symmety graded by vectorial supergenerators (see also \cite{MR1462770}), as well as a case with  $sp(2N)$ singleton bosonic symmetry, accompanied by triplet of spinorial supergenerators (whose bracket relations are no longer pure anticommutators), but with an undeformed  $osp(1|2N)$ subalgebra. A subsequent issue is to establish a relation with (super)Yangian truncations, as has been done for certain classes of finite $W$-algebras and superalgebras 
\cite{MR1682002,MR1965473,MR3147661}. 

The existence of field theoretical realizations of the quadratic supersymmetries as discussed here, perhaps as supersymmetric 
systems with constraints, is of course an open question. While infinite dimensional $W$-algebras arise as spectrum generating algebras in certain lattice models \cite{MR1162083}, and as higher 
spin current algebras in conformal field theories, there exists no concrete
field theoretical construction of $W$-algebras beyond dimension 1 or 2.
It is to be hoped however, that the rich theoretical understandings of $W$-superalgebras and Yangian superalgebras, which can be brought to bear 
on quadratic spacetime supersymmetries with their transcription into this formalism, will be able to inform further progress on these questions.  
 
\subsection*{Acknowledgements:}
ER thanks the SMRI International Visitor program at the University of Sydney, 
and the University of Tasmania, Discipline of Mathematics, for their support and 
their warm hospitality during his visit to Hobart in October 2019.

%

\end{document}